# ESA Science Programme Missions: Contributions and Exploitation – XMM-Newton Observing Time Proposals

**Arvind N. Parmar · Norbert Schartel · Maria Santos-Lleó**

**Abstract** We examine the outcomes of the regular announcements of observing opportunities for ESA's X-ray observatory XMM-Newton issued between 2001 and 2021. We investigate how success rates vary with the lead proposer's gender, "academic age" and the country where the proposer's institute is located. The large number of proposals (10,579) and more than 20 years operational lifetime enable the evolution of community proposing for XMM-Newton to be probed. We determine proposal success rates for high-priority and all proposals using both the numbers of accepted proposals and the amounts of awarded observing time. We find that male lead proposers are between 5–15% more successful than their female counterparts in obtaining XMM-Newton observations. The gender balance and the percentage of successful young proposers are comparable to those of HST after the introduction of dual-anonymous reviewing of HST proposals. We investigate potential correlations between the female-led proposal success rates and the amount of female participation in the Time Allocation Committee. We propose additional investigations to better understand the outcomes presented here.

A. Parmar
Former Head of the Science Support Office
Directorate of Science, ESA, ESTEC
The Netherlands
*Present address:*
Department of Space and Climate Physics
MSSL/UCL
Dorking
UK
E-mail: arvind.parmar@ucl.ac.uk

N. Schartel
Directorate of Science, ESA, ESAC
Villanueva de la Cañada, Madrid, Spain.

M. Santos-Lleó
Directorate of Science, ESA, ESAC
Villanueva de la Cañada, Madrid, Spain.

# 1 Introduction

XMM-Newton was launched on 10 December 1999 into a 48-hour highly elliptical orbit. The nominal mission lifetime was 5 years with a designed lifetime of 10 years. The mission provides sensitive X-ray imaging and spectroscopic observations of a wide variety of cosmic sources from nearby solar system objects to the most distant black holes [1]. The payload consists of the European Photon Imaging Camera (EPIC); [2], Reflection Grating Spectrometer (RGS);[3]) and the Optical Monitor (OM); [4]. The EPIC consists of three imaging spectrometers each located at the focus of an X-ray optic consisting of 58 nested Wolter I geometry mirrors. Two of the EPIC cameras are based on MOS-CCD technology and share the mirrors with RGS grating arrays [5] while the detector based on pn-CCD technology is located behind a fully open telescope [2]. The overall effective area is 2500 cm$^2$ at 1.5 Kilo Electron Volt (keV) and the spatial resolution is 15 arc seconds (half-energy width) and 6 arc seconds (Full-Width at Half Maximum (FWHM)) with a Field of View (FOV) of ∼30 arc minutes diameter. The RGS provides 0.35–2.4 keV spectra with an E/$\Delta$E of 300–700 (1$^{st}$ order). The effective area for the two grating arrays varies in the range of 40–110 cm$^2$ over the energy range. The OM provides optical and Ultra-Violet (UV) monitoring of fluxes through various filters as well as spectroscopy with two grisms. Normally, all three instruments are operated simultaneously.

The scientific products from each observation, produced by a standard pipeline processing, are placed into the XMM-Newton Science Archive (XSA). During any proprietary period data in the XSA are only made available to the Principal Investigator (PI) of the observation, otherwise all the products stored in the XSA are available for public download via the internet. More than 2000 users download data each month. Investigators located in the USA are served from the Goddard Space Flight Center (GSFC) archive which also provides proprietary data for US PIs. General information such as observing date and time, instrument use, the name of PIs etc. can be accessed by the general public for all observations in the XSA. Also, links to the refereed papers that make use of data from each observation can be obtained from the XSA. The key to accessing a data set belonging to a specific observation is the unique 10-digit Observation Identifier (ObsID), consisting of the six-digit proposal number and the four-digit observation identifier.

# 2 Science Results

XMM-Newton is a highly successful observatory with over 6900 refereed publications by the end of 2021. It was launched shortly after National Aeronautics and Space Administration (NASA)'s Chandra X-ray observatory [6]. The two missions are highly complementary with Chandra providing higher spatial resolution imaging whilst XMM-Newton has a larger collecting area. XMM-Newton observes celestial sources ranging from charge-exchange radiation in the upper Earth's magnetopause [7] to the most distant quasars [8], [9], [10] and [11]. Observations with these two missions have changed X-ray astronomy from being a specialised pursuit to one that contributes to nearly all areas of modern observational astrophysics. The scientific impact of XMM-Newton and Chandra is so far reaching that Nature published reviews of their impact in 2009 [12] and again in 2022 [13]. In the following we give some examples that illustrate the scientific impact of XMM-Newton:

**Solar System:** XMM-Newton observations have lead to the detection of charge exchange radiation from Mars [14]. Charge exchange occurs when a neutral atom and an ion interact and an electron is transferred together with the emission of one or more photons.

This was the fist time that this mechanism was identified in the radiation from another planet. XMM-Newton and Chandra found independent pulsations of Jupiter's northern and southern X-ray aurorae [15]. Simultaneous observations by XMM-Newton and the Juno [16] spacecraft revealed the source of Jupiter's auroral flares [17] which are explained through a pulsating magnetic field caused by interactions with the solar wind.

**Exoplanets:** XMM-Newton and Chandra observations discovered the first planet in another galaxy [18]. The study of exoplanets and the interactions with their host stars is an emerging area of X-ray astrophysics. XMM-Newton observation have revealed that the corona of the star HD 189733 flares in phase with the orbit of its hot-Jupiter exoplanet [19]. Helium absorption is an important diagnostic of exoplanet atmospheres, particularly in hot Neptune and Jupiter systems which may have thick evapourating atmospheres producing gaseous tails streaming behind the planet. XMM-Newton observations have revealed the dependence of the Helium absorption on a sample of host stars' X-ray luminosities [20] allowing an improved characterisation of the atmospheres and contributing physical processes to be made.

**Millisecond Pulsars:** XMM-Newton observations were key in identifying a millisecond pulsar which cycles between emitting radio and X-rays. The object is the evolutionary link between accretion and rotation-powered millisecond pulsars. The existence of such objects had been predicted, but never found [21]. Soft Gamma-ray Repeaters are slowly rotating isolated neutron stars that sporadically undergo outbursts. These are assumed to be powered by their internal magnetic energy with dipole field strengths of $10^{14} - 10^{15}$ Gauss and are often called "magnetars". A surprise was the detection of a Soft Gamma-ray Repeater, SGR 0418+5729, with a low-magnetic-field strength as determined from its timing properties [22]. XMM-Newton detected a rotation-phase dependent proton cyclotron absorption feature from this source, which can be interpreted as originating from a multi-pole component of the magnetic field which has a strength in the range of expectation for magnetars [23]. An ultra-luminous X-ray source is an exceptionally bright X-ray source that is not coincident with its galactic nucleus. The detection of pulsations from some of these sources indicates a neutron star origin [24], [25] and therefore excludes intermediate-mass black holes for at least part of the population.

**Supermassive Black Holes:** XMM-Newton has been used to investigate the environments close to supermassive black holes using a technique called reverberation mapping. This utilises variations in the X-ray continuum emission from the corona and the X-ray line emission originating in the inner regions of the accretion disk which "reverberate" with a time delay due to the light travel time [26], see also [27]. The current state of the art was demonstrated by XMM-Newton observations of the highly variable Active Galactic Nucleus (AGN) IRAS 13244-3809 [28]. The XMM-Newton campaign allowed the inherent degeneracy between the black hole mass, inner disk radius and height of the X-ray emitting corona to be broken. This allowed, for the first time, the simultaneous determination of the supermassive black hole mass and spin. A flare during the observation of the AGN 1 Zwicky 1 allowed the light bending and X-ray echos from behind a supermassive black hole [29] to be traced. XMM-Newton was of fundamental importance in the detection and exploration of fast and ultra-fast outflows of highly ionized matter from the nuclei of AGNs [30], [31], [32], [33], [34]. The growth of galaxies and cluster of galaxies is significantly slower over cosmological time scales than predicted by simulation. Outflows from AGNs are a strong candidate explain this lower growth through feedback on the cosmological structure formation.

**Tidal Disruption Events:** XMM-Newton has been used to study the Tidal Disruption Event (TDE)s of stars captured by black holes. This allows the properties of black holes in the centre of normal galaxies to be investigated [35]. Combining observations from XMM-Newton and Chandra a TDE from an elusive intermediate-mass black hole in a star cluster was detected [36] and a decade long TDE was traced [37]. Quasi-periodic oscillations were detected from two TDEs. These probably originate from areas close to the event horizons of the black holes [38], [39]. Blue-shifted absorption lines of highly ionized atoms may also be explained though emission close to the black hole's event horizon [40]. Even relativistic reverberation in the accretion flow of one TDE was evident based on XMM-Newton data [41]. XMM-Newton leads in the detection and exploration of quasi-periodic eruptions[42], [43], which are most likely explained by localised disruptions of accretion disks around supermassive black holes by orbiting compact objects following (partial) tidal disruption events.

**Cluster of Galaxies:** XMM-Newton observations have allowed a universal mass profile for clusters of galaxies to be established [44] as well as determining the origin of the metals observed in them [45],[46]. Some 30% of the metals can be traced to Type I supernovae and 70% to core collapse supernovae. One of the fist achievement of XMM-Newton was the non-detection of strong X-ray emission from cooling flows in the centre of clusters of galaxies as was generally expected [47] [48], [49]. The low upper-limits on turbulence induced broadening of emission lines in the XMM-Newton RGS spectra of some of the brightest clusters of galaxies, indicates the the dissipation of turbulence may prevent cooling of the cluster core [50], [51].

**Cosmology:** An RGS observation of a distant AGN has allowed the detection of the Warm-Hot Intergalactic Medium (WHIM) for the first time [52] closing the gap between the number of observed baryons and number of baryons predicted by big-bang nucleosynthesis. By combining gravitational lensing observations of the total mass with optical and infrared observations of the cold baryonic mass and XMM-Newton observations of the hot baryonic mass in the WHIM, the dark matter large-scale distribution in the Cosmos field has been determined [53]. It is found to be consistent with the predictions of cosmic structure formation. XMM-Newton plays an important role in the search for dark matter particles [54], [55], [56]. An analysis of ~450 clusters observed by XMM-Newton in ~50 square degrees of the extra-galactic sky shows clear tensions with the predictions using the cosmological parameters as determined by Planck 2015 [57], [58]. An analysis of the XMM-Newton follow-up observations of Planck detected clusters of galaxies [59] also reveals tensions with the predictions using the standard concordance cosmological model, as does an analysis of the Hubble Diagram for quasars [60].

## 3 Scientific Productivity

The scientific productivity of XMM-Newton was assessed in 2014 by [61]. For comparison, similar analyses for NASA's Chandra X-ray observatory are to be found in [62] and for Hubble Space Telescope (HST) in [63] as well as for a range of astronomical facilities in [64], [65], [66] and [67]. The ObsID and instruments used to provide the scientific results reported in each publication were determined by [61]. The ObsIDs were then used to access the XSA to provide detailed information on the observations themselves and on the original proposal. The information obtained from these sources was then combined to allow the scientific productivity of the mission to be investigated. The major conclusions of [61] were:

- Around 100 scientists per year become lead authors for the first time on a refereed paper which directly uses XMM-Newton data.

- Each refereed XMM-Newton publication receives an average of around four citations per year with a long-term citation rate of three citations per year, more than five years after publication.
- About half of the articles citing XMM-Newton articles are not primarily X-ray observational papers.
- The distribution of elapsed time between observations taken under the Guest Observer program and first article peaks at 2 years with a possible second peak at 3.25 years. Observations taken under the Target of Opportunity (ToO) program (see below) are published significantly faster, after one year on average.
- The fraction of science time used in at least one publication reaches >95% after 7 years and about 70% after exclusion of catalog-type publications.
- The scientific productivity of XMM-Newton measured by the publication rate, number of new authors and citation rate, remained extremely high with no evidence that it was decreasing after more than (the then) 12 years of operations.

Up to the end of 2021, there have been over 6900 XMM-Newton refereed publications that make direct use of data from XMM-Newton including from its primary catalogues, or make quantitative predictions of results from the mission or describe XMM-Newton, its instruments, operations, software or calibrations.

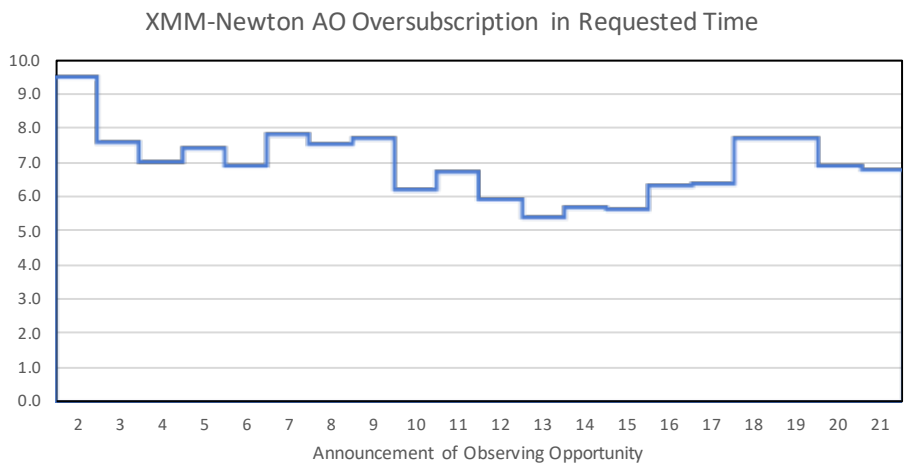

**Fig. 1** The over-subscription of requested XMM-Newton observing time compared to that available for AO-2 to AO-21, covering an interval of 20 years. The values for AO-17 and AO-20 exclude the multi-year heritage programme opportunities which were offered during these AOs.

## 4 Observing Time

The annual calls for observing proposals are for "Open Time" targets. There are four types of XMM-Newton observing time:

1. Open Time: ~90% of the observing time is available for Guest Observers via Announcement of Opportunity (AO)s which are open to scientists worldwide. Open Time includes an-ticipated ToOs – astronomical events observable by XMM-Newton, which cannot be predicted and scheduled at the time of proposal submission.
2. Discretionary Time: The XMM-Newton Project Scientist can grant "Discretionary Time" which is ~5% of the available observing time. This can be used for rapid unanticipated ToOs – observations where a timely observation outside the normal AO cycle is likely to result in a significant scientific impact. Discretionary Time observations may have a proprietary period of six months, or may be made publicly available as soon as the relevant data files have been created.
3. Calibration Time: ~5% of the available observing time is used to maintain the calibration and monitor instrument health.
4. Guaranteed Time: During the first two years of the mission, the scientific groups that provided the instruments and that were involved in the scientific ground segment development were awarded Guaranteed Time observations.

Open Time targets are selected competitively through peer review by an Observation Time Allocation Committee (OTAC). Calls for observing proposals are usually issued annually and are highly oversubscribed (by at least a factor of 5) compared to the available observing time (see Fig. 1). Proposals are submitted electronically to the XMM-Newton Science Operations Centre (SOC). Following receipt proposals are submitted to the OTAC for scientific review while SOC staff perform visibility assessments and duplication checks. The feasibility assessment is performed by panel members of the OTAC. The PIs of successful proposals are required to provide detailed information on the requested configurations in a second submission phase. The PIs of successful proposals are granted a proprietary period of one year before the data are made publicly available. The types of Open Time proposals available has evolved as the mission's science has matured and as new opportunities to collaborate have become available. For AO-20, these included:

1. Guest Observer. These are "normal" observations including time constrained and observations coordinated with other facilities.
2. Anticipated ToOs.
3. Large Programmes requiring a significant amount of observing time (>300 ksec).
4. Multi-Year Heritage Programmes. These are scientific visionary programmes which need between 2 and 6 Msec of observing time and are performed over three AOs.
5. Fulfil Programmes: To assist in the completion of samples of objects.
6. Joint observations with INTEGRAL, Chandra, and the European Southern Observatory (ESO) Very Large Telescope (VLT), HST, Swift, NuSTAR, High Energy Stereoscopic System (HESS), MAGIC and NRAO/GBO observatories.

With the introduction of Multi-Year Heritage Programmes in 2017, up to four primary investigators are allowed for large programs (>300 ksec) in order to promote cooperation between diverse groups and to support raising the funding necessary for extended data analysis. In these cases, the corresponding author is assumed to be the PI.

## 5 Observation Time Allocation Committee (OTAC)

The OTAC reviews all submitted proposals and makes recommendations on the targets to be observed by XMM-Newton. It takes into account the scientific case, justification, merit and relevance of the proposed observation(s) and the potential contribution of the overall scientific return of the mission. In addition, potential duplication with performed and planned observations with XMM-Newton and Chandra, as well as the technical feasibility, are also considered. The OTAC recommendations on the observing programme are given priority assignments of A, B and C.

Table 1 XMM-Newton AO summary. The numbers of OTAC members include panel chairs.

| AO | Year Issued | No. of Proposals | Over-subscription (Time) | OTAC Members Male | OTAC Members Female | OTAC Panel Chairs Male | OTAC Panel Chairs Female |
|---|---|---|---|---|---|---|---|
| 2 | 2001 | 869 | 9.5 | 41 | 6 | 13 | 2 |
| 3 | 2003 | 692 | 7.6 | 59 | 11 | 12 | 2 |
| 4 | 2004 | 660 | 7.0 | 58 | 6 | 12 | 1 |
| 5 | 2005 | 641 | 7.4 | 56 | 9 | 13 | 0 |
| 6 | 2006 | 594 | 6.9 | 56 | 9 | 13 | 0 |
| 7 | 2007 | 586 | 7.8 | 59 | 6 | 12 | 1 |
| 8 | 2008 | 555 | 7.5 | 52 | 13 | 10 | 3 |
| 9 | 2009 | 539 | 7.7 | 53 | 12 | 12 | 1 |
| 10 | 2010 | 491 | 6.2 | 53 | 17 | 13 | 1 |
| 11 | 2011 | 501 | 6.7 | 57 | 13 | 12 | 2 |
| 12 | 2012 | 475 | 5.9 | 53 | 12 | 12 | 1 |
| 13 | 2013 | 452 | 5.4 | 50 | 15 | 11 | 2 |
| 14 | 2014 | 431 | 5.7 | 53 | 12 | 12 | 1 |
| 15 | 2015 | 431 | 5.6 | 53 | 12 | 10 | 3 |
| 16 | 2016 | 442 | 6.3 | 46 | 19 | 10 | 3 |
| 17 | 2017 | 441 | 6.4 | 51 | 15 | 8 | 5 |
| 18 | 2018 | 442 | 7.7 | 48 | 14 | 7 | 6 |
| 19 | 2019 | 462 | 7.7 | 54 | 11 | 8 | 5 |
| 20 | 2020 | 447 | 6.9 | 47 | 16 | 7 | 5 |
| 21 | 2021 | 428 | 6.8 | 33 | 22 | 6 | 5 |
| **Total** | | **10579** | | **1032** | **250** | **213** | **49** |

- Priority A – The highest priority targets (about 10% of the available observing time) that are of major scientific importance and are scheduled with the highest priority. Observations that are not completed within the observing period are automatically transferred to the next one.
- Priority B – As per Priority A, but for the next highest 40% of observing time.
- Priority C – Targets that are used as "fillers" for around 50% of the observing time and have a significantly lower chance of being observed. Targets that have not been observed by the end of the observing period are not transferred to the next one.

An average of 530 valid proposals were received in response to each AO (see Table 1). With an over-subscription factor that remains high during all the AOs (see Fig. 1 and Table 1). The scientific assessment of the proposals is a major undertaking by the astronomical community with a significant fraction of the high-energy community having participated so far. The OTAC is divided into different scientific categories to reflect the range of topics proposed. These categories have changed over time reflecting the scientific evolution and consequently different number of requests in the various categories. Currently the categories include stars and planets, compact objects, active galactic nuclei, individual, groups and cluster of galaxies, and cosmology. Much of the assessment is performed by panels, consisting of five scientists selected from the worldwide community. The majority of OTAC members are selected from the participants of previous AOs, i.e. scientists who proposed for XMM-Newton time are often invited to be members of subsequent OTACs. Many categories require multiple panels and there are typically between 11 and 15 panels for each AO including dedicated panels for Multi-year Heritage Programmes. Each panel is led by a panel chair and the overall OTAC is led by the chairperson. The names of the OTAC members and chairs (with exception of the chairperson who sits on the XMM-Newton Users' Group (XUG), see Sect. 6) are not public.

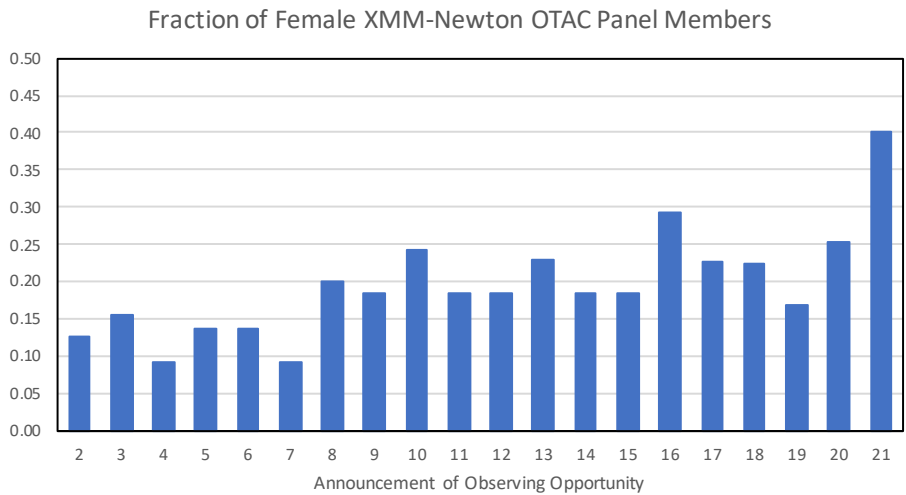

**Fig. 2** The fraction of female XMM-Newton OTAC members compared to the total for AO-2 to AO-21 covering an interval of around 20 years. A steady increase in the fraction of females from ∼0.15 to ∼0.25 by AO-20 is evident with a sharp increase in AO-21.

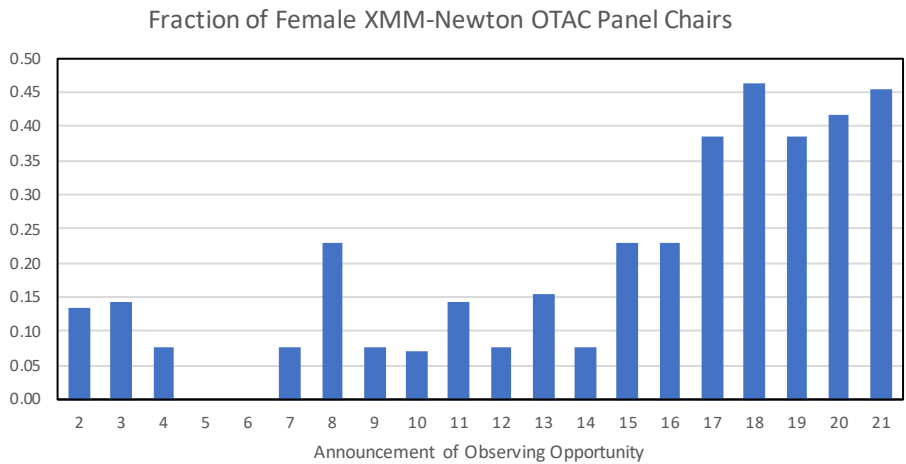

**Fig. 3** The fraction of female XMM-Newton OTAC chairs compared to the total for AO-2 to AO-21 covering an interval of around 20 years. An increase in the fraction of females from ∼0.10 to ∼0.45 by AO-21 is evident.

The genders of the OTAC members and panel chairs were assigned through the personal knowledge of the authors and SOC staff. We appreciate that gender identity is more complex than a binary issue. However, no attempt was made to assign genders other than male or female in this study. We then examined the gender composition of the OTAC members and panel chairs (Table 1) over the 20 years between AO-2 to AO-21. AO-1 was opened prior to the XMM-Newton launch and used different software which is no longer maintained. This means that some of the information needed for comparisons with later AOs is not available. For this reason only the results of AO-2 to AO-21 are considered here.

OTAC panel members are chosen for their high-level of relevant scientific knowledge, while selection of the OTAC chairs is for scientists considered to have leadership roles in high-energy astronomy. Fig. 2 shows the fraction of female XMM-Newton OTAC members from AO-2 to AO-21 and Fig. 3 the same for the panel chairs. In total there were 1032 male and 250 female panel members and 213 male and 49 female panel chairs. Since members and chairs may serve for multiple AOs, the numbers of individuals involved is smaller. Most of the OTAC members (and chairs) are based at institutes located in the European Space Agency (ESA) Member States with a significant number from institutes in the United States. An overall increase in the fraction of females from ∼0.15 to ∼0.25 by AO-20 is evident with a sharp increase in AO-21. There is a marked increase in the fraction of females in leadership positions as panel chairs from ∼0.1 of the total number of chairs to ∼0.45 by AO-21.

## 6 XMM-Newton User Group

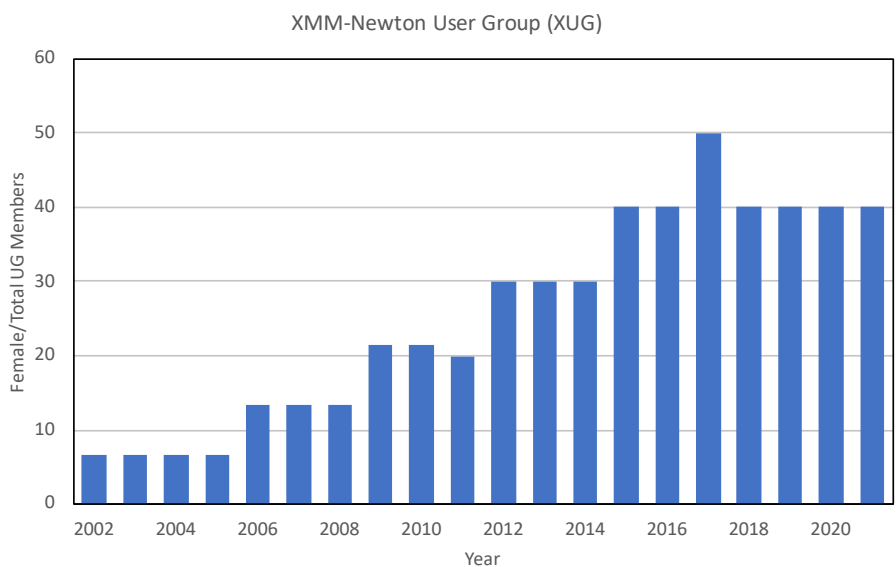

**Fig. 4** The percentage of female XMM-Newton XUG members compared to the total between 2002 and 2021. An increase in the percentage of female members from ∼5% to ∼40% of the total is evident.

As well as the OTAC, there is another body involved in optimising the scientific output of the mission. This is the XUG which advises ESA, through the Project Scientist, on all matters relating to the optimisation of the scientific output of the mission. It also acts as a forum to discuss input from the community of users, and when appropriate advise or recommend action to ESA regarding XMM-Newton operations. There were 15 members (including the chair) when the XUG was established in 2002 which decreased to ten members in 2011. The chair of the OTAC is also a member of the XUG to ensure good coordination between the two bodies. In addition, the XMM-Newton Project Scientist, Mission Manager, Science Support Manager and instrument principal investigators attend parts of the meetings. The names of all the members are made public. Genders were assigned in the same way as for the OTAC. Fig. 4 shows the number of female XUG members divided by the total between 2002 and 2021. An notable increase in the fraction of female members from ∼5% to ∼40% of the total is evident.

## 7 Proposers

Between AO-2 and AO-21 there were 10,579 proposals submitted in response to XMM-Newton calls for observing opportunities. This provides a rich data set to quantify many aspects of the proposal process. When submitting a proposal, XMM-Newton proposers are required to indicate the country where their institute or university is located using a drop-down menu. This information allows the nationalities of the proposers' institutes to be examined. Proposers are not required to submit information on gender, age or type of position held and these have to be derived by other means if required.

### 7.1 Proposers' Institute Countries

Table 2 shows the number of proposals submitted and accepted for a range of countries including all 22 ESA Member States and other countries from which significant numbers of proposals originated. Accepted proposals are ones which were awarded any observing time. Only 31 proposals were received from countries that are not listed individually in Table 2.

Scientists located at institutes within the United States submitted the most proposals – 40.7% of the total. This is followed by Germany (11.9%), Italy (11.5%), the United Kingdom (8.9%), Spain (5.7%) and France (3.7%). Figure 5 shows the variations in percentage of proposal numbers versus AO for six ESA Member States (Germany, France, Italy, the Netherlands, Spain and the United Kingdom). These show relatively stable proposal submission fractions with AO number with the possible exception of the United Kingdom which shows a decline from ∼15% for the earliest AOs to ∼6% for the latest. There is evidence for a small increase in the submission fraction with time of proposers located in Germany.

Figure 6 shows the variations in percentage of proposal numbers versus AO for four Non-Member States (Canada, China, India and Japan). China shows an increase of the proposal submission fractions with AO number from 0% to ∼3%. A similar, but weaker, trend is seen for India where the submission fractions increases from 0% to ∼1.5%. The submission fractions with AO number of Japan shows a steep decline from ∼7% for AO-2 to ∼2% in AO-4 and stays stable at this level for all later AOs. The decline may reflect the successful launch of the Japanese Aerospace Exploration Agency (JAXA)/NASA X-ray mission Suzaku and its scientific exploitation in the subsequent years.

In order to be able to draw reliable conclusions, we examined the number of accepted proposals for the 13 countries whose scientists have submitted >80 proposals (Table 2). The sum of all the submitted proposals from these countries is 95% of the total number of submitted proposals. The countries with the highest accepted fractions are the Netherlands where 51.2% of 361 proposals were accepted, followed by France with 48.0% of 394 proposals, Italy with 44.3% of 1215 proposals, Switzerland with 43.5% of 115 proposals, the United Kingdom with 43.1% of 941 proposals and Germany with 42.5% of 1258 proposals. This may reflect the well-established high-energy communities in these countries.

Table 2 The number of proposals submitted and accepted from PIs located in a range of countries. In this case "accepted" means awarded any observing time.

| Country | No. Proposals Submitted | Percentage of Total | No. Accepted | Acceptance Percentage |
|---|---|---|---|---|
| AUSTRIA | 26 | 0.25 | 6 | 23.1 |
| BELGIUM | 150 | 1.42 | 57 | 38.0 |
| CZECH REPUBLIC | 15 | 0.14 | 5 | 33.3 |
| DENMARK | 9 | 0.09 | 5 | 55.6 |
| ESTONIA | 4 | 0.04 | 0 | 0.0 |
| FINLAND | 49 | 0.46 | 16 | 32.7 |
| FRANCE | 394 | 3.72 | 189 | 48.0 |
| GERMANY | 1258 | 11.89 | 535 | 42.5 |
| GREECE | 40 | 0.38 | 12 | 30.0 |
| HUNGARY | 6 | 0.06 | 0 | 0.0 |
| IRELAND | 32 | 0.30 | 6 | 18.8 |
| ITALY | 1215 | 11.49 | 538 | 44.3 |
| LUXEMBOURG | 0 | 0.00 | 0 | 0.0 |
| NETHERLANDS | 361 | 3.41 | 185 | 51.2 |
| NORWAY | 1 | 0.01 | 1 | 100.0 |
| POLAND | 34 | 0.32 | 12 | 35.3 |
| PORTUGAL | 4 | 0.04 | 1 | 25.0 |
| ROMANIA | 0 | 0.00 | 0 | 0.0 |
| SPAIN | 598 | 5.65 | 238 | 39.8 |
| SWEDEN | 17 | 0.16 | 6 | 35.3 |
| SWITZERLAND | 115 | 1.09 | 50 | 43.5 |
| UNITED KINGDOM | 941 | 8.89 | 406 | 43.1 |
| ARGENTINA | 14 | 0.13 | 5 | 35.7 |
| AUSTRALIA | 43 | 0.41 | 17 | 39.5 |
| BRAZIL | 18 | 0.17 | 10 | 55.6 |
| BULGARIA | 13 | 0.12 | 6 | 46.2 |
| CANADA | 161 | 1.52 | 65 | 40.4 |
| CHILE | 35 | 0.33 | 12 | 34.3 |
| CHINA | 116 | 1.10 | 31 | 26.7 |
| INDIA | 83 | 0.78 | 20 | 24.1 |
| ISRAEL | 28 | 0.26 | 12 | 42.9 |
| JAPAN | 304 | 2.87 | 90 | 29.6 |
| KOREA | 20 | 0.19 | 5 | 25.0 |
| MEXICO | 35 | 0.33 | 9 | 25.7 |
| RUSSIA | 28 | 0.26 | 9 | 32.1 |
| TAIWAN | 46 | 0.43 | 10 | 21.7 |
| TURKEY | 29 | 0.27 | 10 | 34.5 |
| UNITED STATES | 4306 | 40.70 | 1699 | 39.5 |
| OTHER | 31 | 0.29 | 9 | 29.0 |

### 7.2 Proposers' Gender

We next examined the number of proposals with male and female PIs from these 13 countries (Table 3). Since proposers are not asked to specify their gender (nor age or type of position held etc.) gender information for each proposer was obtained through examining publicly-accessible web-based data in a similar way as for the HST study by I.N. Reid [68]. We appreciate that gender identity is more complex than a binary issue, however, no attempt can be made to assign genders other than male or female as this information is not readily available.

**Table 3** The number of proposals submitted and accepted from male and female PIs located in countries with >80 proposals in total. The differences are with respect to the acceptance percentages given in Table 2 which are for all proposals.

| Country | Proposals Submitted | | % Female | Proposals Accepted | | % Difference | |
|---|---|---|---|---|---|---|---|
| | Male PI | Fem. PI | /Total | Male PI | Fem. PI | Male PI | Fem. PI |
| BELGIUM | 86 | 64 | 42.7 | 35 | 22 | 2.7 | −3.6 |
| FRANCE | 258 | 136 | 34.5 | 141 | 48 | 6.7 | −12.7 |
| GERMANY | 945 | 313 | 24.9 | 413 | 122 | 1.2 | −3.6 |
| ITALY | 843 | 372 | 30.6 | 389 | 149 | 1.9 | −4.2 |
| NETHERLANDS | 267 | 94 | 26.0 | 130 | 55 | −2.6 | 7.3 |
| SPAIN | 431 | 167 | 27.9 | 175 | 63 | 0.8 | −2.1 |
| SWITZERLAND | 96 | 19 | 16.5 | 45 | 5 | 3.4 | −17.2 |
| UNITED KINGDOM | 752 | 189 | 20.1 | 322 | 84 | −0.3 | 1.3 |
| CANADA | 80 | 81 | 50.3 | 33 | 32 | 0.9 | −0.9 |
| CHINA | 103 | 13 | 11.2 | 31 | 0 | 3.4 | −26.7 |
| INDIA | 60 | 23 | 27.7 | 17 | 3 | 4.2 | −11.1 |
| JAPAN | 234 | 70 | 23.0 | 66 | 24 | −1.4 | 4.7 |
| UNITED STATES | 3400 | 906 | 21.0 | 1374 | 325 | 1.0 | −3.6 |

Among the countries with more than 80 proposals submitted, the country with the highest fraction of female PIs is Canada (50.3%) followed by Belgium (42.7%). The countries with the lowest fraction of female PIs are China (11.2%) and Switzerland (16.5%). Figure 7 shows the acceptance fraction differences for male and female PIs compared to the average for that country for each of the 12 countries with >100 proposals. The two countries with the lowest fraction of submitted proposals from female PIs (China and Switzerland) also have the lowest fractions of proposals with female PIs accepted of −26.7% and −17.2%, respectively, compared to the average. Only three countries have better female PI acceptance rates than for male PIs – the Netherlands, Japan and the United Kingdom with 7.3%, 4.7% and 1.3%, respectively compared to the averages.

Figure 8 shows the fraction of XMM-Newton proposals with female PIs compared to the total. This increased relatively steadily from ∼20% to ∼30% between AO-2 and AO-21 (2001 to 2021).

### 7.3 Proposers' Academic Age

In order to determine the "academic age" of proposers we used the difference between the year that their Doctor of Philosophy Degree (PhD) was awarded and the year that an AO to which they submitted a proposal was issued. Thus a proposer who submits multiple proposals to the same AO will be counted multiple times in the same "academic year", whilst one who submits multiple proposals to different AOs will be counted in different "academic years".

The year a proposer obtained their PhD (or equivalent) was determined for 94.8% of the proposals by searching the internet, particularly sites such as the Astrophysics Data Service (ADS), LinkedIn, the Astronomy Genealogy Project (astrogen.aas.org), ORCID.org,

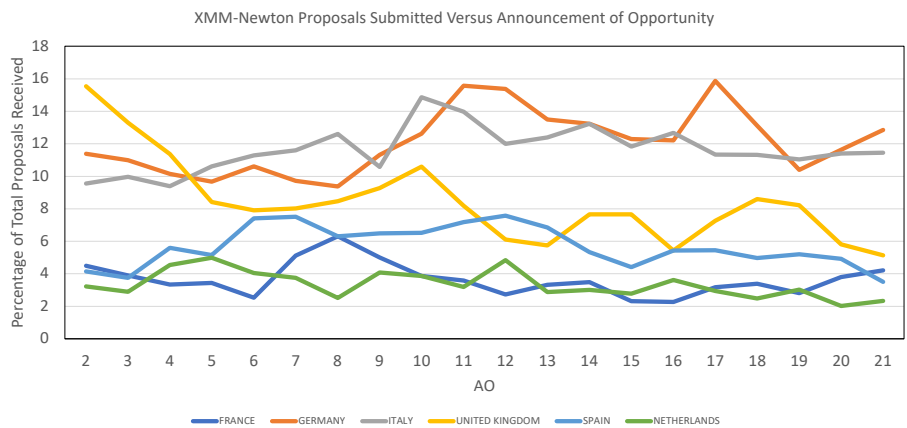

**Fig. 5** Percentages of proposals submitted from the six ESA Member States with the largest numbers of proposals between AO-2 to AO-21. A decline in the percentage of proposals from the United Kingdom from ~15% for the earliest AOs to ~6% for the latest is evident.

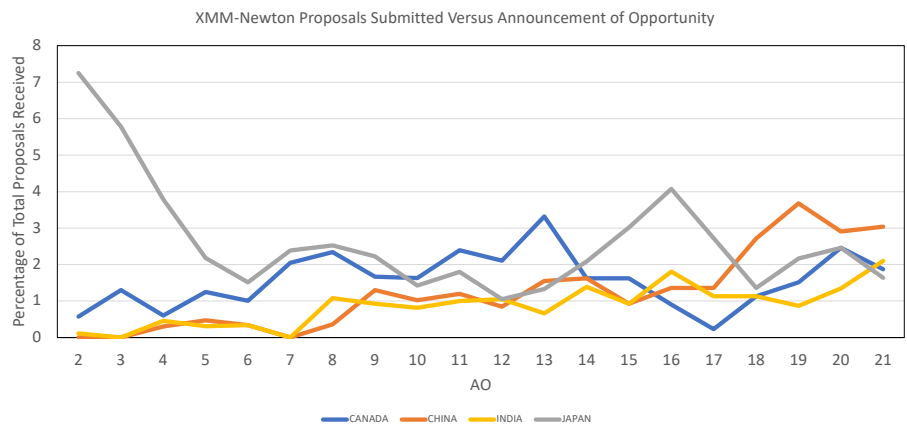

**Fig. 6** Percentages of proposals submitted from the Non-Member States with the largest numbers of proposals (excluding the USA) between AO-2 to AO-21. China and India show increases with AO number from 0% to ~3% and from 0% to ~1.5%, respectively. The large decrease in the percentage of proposals from Japan may be related to the launch of the JAXA/NASA Suzaku X-ray mission and its subsequent exploitation.

IEEE Xplore (`https://ieeexplore.ieee.org/Xplore/home.jsp`), and, for French theses `https://www.theses.fr`. Some proposers were contacted directly and provided their PhD dates by email. The proposers for which the PhD could not be found are often retired, deceased or have left astronomy. For PhD students who had not yet completed their degrees, the expected year of submission was used. For the small number of proposers who did not have a PhD and were not enrolled in a PhD programme, their dates were assumed to be arbitrarily far in the future. For some late career scientists in Italian institutes who do not have a PhD, their "academic age" was taken to be three years after they obtained their Laurea. It should be noted that using the year of PhD to indicate the number of years experience neglects time spent outside of astronomy. We note that the "youngest" proposer was 13 years before obtaining a PhD and the oldest 52 years after! The mean "academic age" of PIs is 10.9 years. For female PIs it is 9.8 years post-PhD, compared to 11.2 years for male PIs.

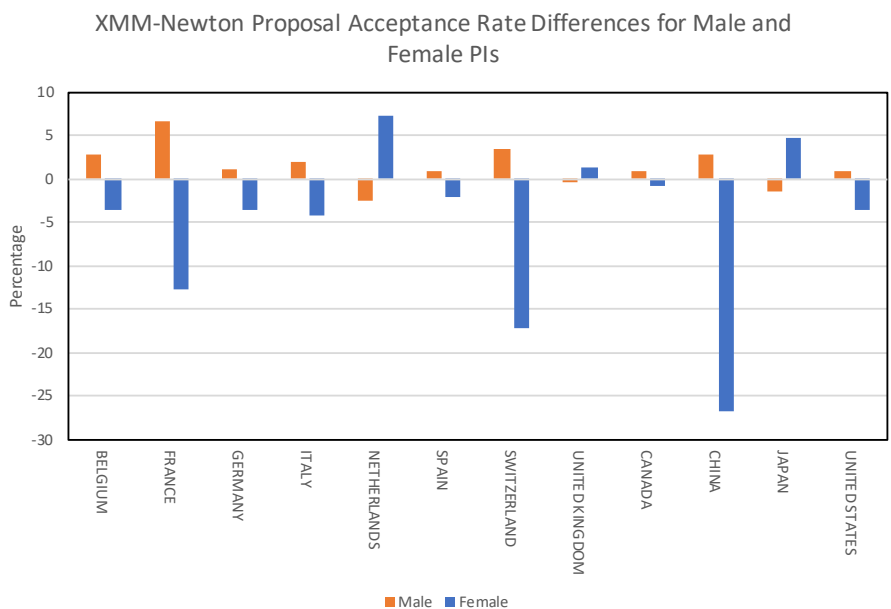

**Fig. 7** The acceptance fraction differences of XMM-Newton proposals with male and female PIs compared to the average for all proposers from that country. The countries are those with >100 submitted XMM-Newton proposals.

## 8 Proposal Selection

We evaluated the outcomes of the XMM-Newton AO-2 to AO-21 selection processes for both genders using two different methods:

1. Proposals that were awarded priority A, B or C observing time.
2. Proposals that were awarded high-priority (A or B) observing time.

Observations from high-priority proposals (A or B) are guaranteed to be performed, so having an approved high-priority proposal is more likely to result in scientific papers and so be beneficial for the career of a researcher. In contrast only around 40% of priority C proposals are actually observed. Since many XMM-Newton proposals are allocated less observing time than requested, almost always because not all the requested targets are approved, we decided to examine both the numbers of successful proposals and how much observing time was awarded. We note that in the later AOs there was significant time allocated to Large and Multi-Year Heritage Programmes that required substantial investments in observing time. If the allocation of observing time is handled differently for males and female PIs, then this will show as a difference in the relative amounts of time approved, compared to the number of proposals accepted. Thus, for both outcomes, the numbers and amounts of awarded observing times were determined for the total, male PI and female PI populations.

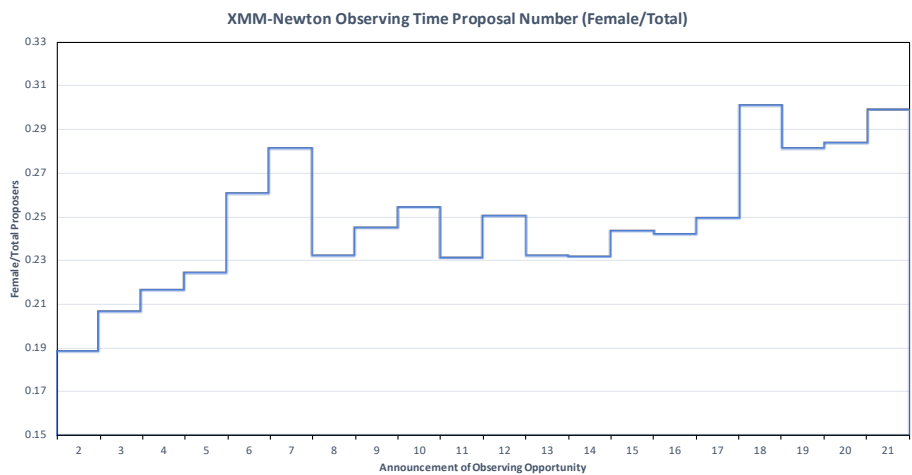

**Fig. 8** The fraction of submitted proposals with a female PI compared to the total between AO-2 to AO-21 (2001 to 2021). Results are for all countries where PIs submitted proposals.

### 8.1 Proposal Selection – Priority A, B or C

We first examined the number of proposals that were awarded *any* observing time. A total of 10,579 proposals were submitted to XMM-Newton AO-2 to AO-21 of which 4287 were awarded observing time. This corresponds to an overall success rate of 40.5%. A total of 7997 proposals with male PIs and 2582 proposals with female PIs were submitted, of which 3316 and 971 proposals were awarded *any* observing time in priority A, B, or C. This gives success rates of 41.5% and 37.7% for male and female PI proposals, respectively. This is a difference in favour of males of 10.1%.

Proposal submissions and proposal acceptance are unlikely to be independent or random processes and for Poisson statistics to apply events need to be independent of each other. However, to illustrate the uncertainties that would apply if such statistics are applicable, we used square root uncertainties to find success rates of $(41.5 \pm 1.0)\%$ and $(37.7 \pm 1.8)\%$ for male and female PI proposals, respectively. This is a difference in favour of males of $(10.1 \pm 5.9)\%$ which is formally significant at $1.7\sigma$. We emphasise that these uncertainties are only given to illustrate the outcomes if Poisson statistics were to apply to the selection process. We note that actual uncertainties on the proposal numbers for each AO are zero. We further note that there are systematic uncertainties associated with this process due to misappropriated genders and incorrect PhD "academic ages", but these are likely to be too small to significantly affect our calculations.

The success rates are shown for male, female and all proposers for each AO in Fig. 13. In contrast to the results from HST Cycles 11 to 20 reported in [68] where proposals with male PIs had a consistently higher success rate than those with female PIs, XMM-Newton proposals with male PIs had higher success rates for 14 of the AOs and female PIs had higher success rates for 5 AOs. For AO-21, the success rates are almost identical. Interestingly, there is no obvious evolution in the male and female PI proposal acceptance rates with AO number over an interval of more than 20 years.

Figure 14 shows the variation in acceptance rates more clearly. For each AO, it shows the difference between the expected number of accepted proposals, calculated using the overall acceptance rate, and the number actually accepted. It shows that the largest discrepancies occurred during AO-6 and AO-14 where proposals with a female PIs had lower success rates of 16.4% and 10.6%, respectively, compared to proposals with a male PI. The most successful AO for female PI proposals was AO-12 where female-led proposals were 5.4% more likely to be accepted than ones with a male PI.

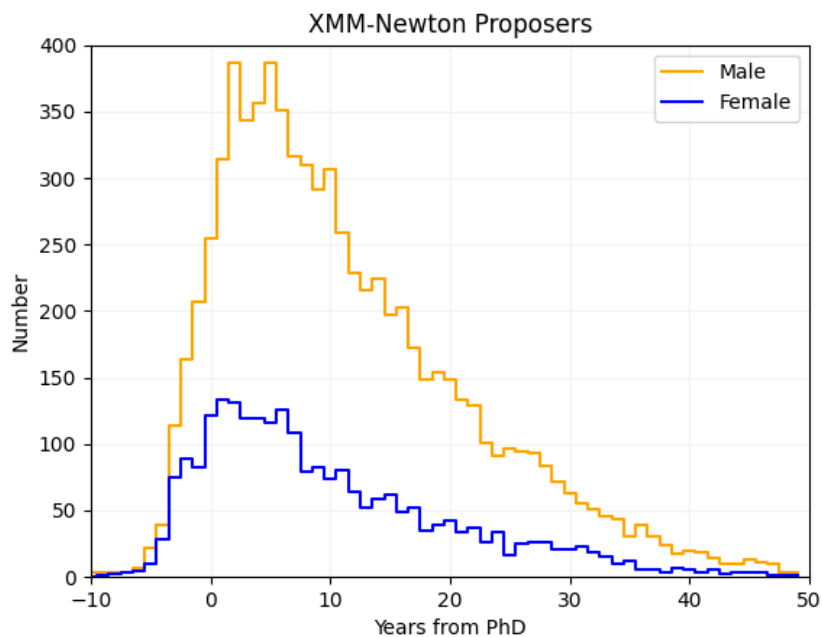

**Fig. 9** The "academic age" distribution (years since PhD) for male and female XMM-Newton PIs between −10 and 50 years. The mean "academic age" PIs is 10.9 years. For female PIs this is 9.8 years post-PhD, compared to 11.2 years for male PIs.

We examined the OTAC membership genders to see if these are correlated with proposal acceptance rates. Figures 16 and 17 show the acceptance fraction of priority A and B proposals with female PIs against the fractions of female compared to the total number of OTAC members and the fraction of female OTAC panel chairs, respectively. It can be seen that there is no obvious correlation between the success rates of female-led proposals and the fraction of female OTAC members or panel chairs.

The female PI success rate during AO-3 of 0.44 was unusually high, but based on only 63 proposals accepted with Priority A or B. There were no female panel chairs for AO-5 and AO-6. AO-5 has an average female PI Priority A and B success rate of 0.23, whereas AO-6 has the lowest female PI success rate of any of the AOs of 0.12 (Fig. 3). This supports the view that having females in leadership positions in the OTAC may improve female PI proposal success rates.

Figure 15 shows the same excesses normalised by dividing by the square root of the expected number of accepted proposals. As discussed earlier, Poisson statistics do not necessarily apply to the peer-review process. Indeed, if the individual results are examined, the individual results could arise from sampling statistics of an underlying distribution with equal male and female success rates. However, such an argument is more difficult to sustain when viewed in the context of the distribution of results from all 20 AOs.

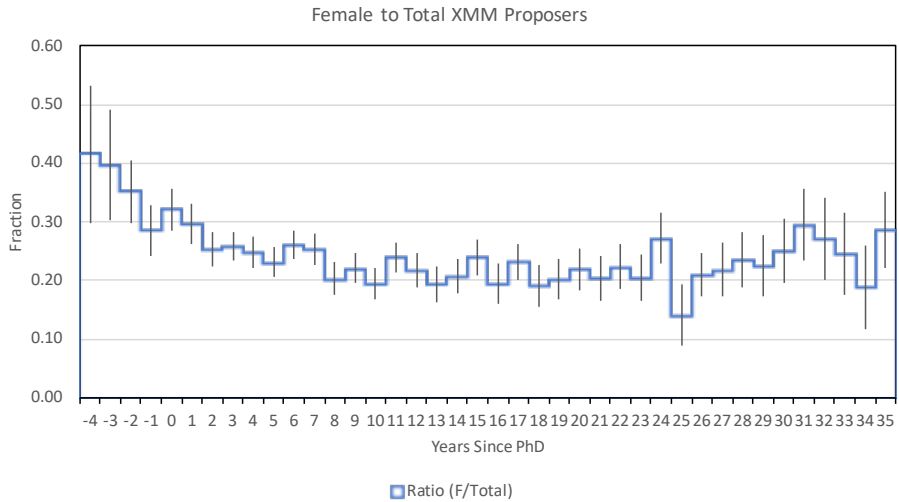

**Fig. 10** The ratio of the "academic age" (years since PhD) of female compared to all XMM-Newton PIs showing the increase in early career (less than around 5 years from obtaining a PhD) female PIs compared to the total. The indicative error bars show $1\sigma$ standard deviations assuming that the number of proposers in each age bin follows a Poisson distribution.

We next examined the observing times awarded in priority A, B and C. The average success rate for time is 24.7% for male PIs and 23.6% for female PIs of that requested (Table 7). This is a difference of 4.5% in favour of male PIs. This smaller difference when the number of proposals are compared of 10.1% and may suggest that female-led proposers are awarded relatively more observing time than their male counterparts.

These results indicate that proposals with male PIs are more likely to be approved than those with female PIs. This prompted us to examine whether differences in the population of XMM-Newton male and female observing time proposers could account for this difference. This could happen for example if the underlying male population is more experienced and if the proposal success rate increases with experience as would seem likely.

Figure 9 shows the numbers of male and female XMM-Newton PIs who obtained their PhDs or equivalent in five year bins between 4 years before the PhD date to 35 years after. A two-sample Anderson-Darling Test (e.g., [69]) shows that the hypothesis that both samples come from the same underlying population can be rejected at >99% confidence. The differences between the two samples can be more clearly seen in Fig. 10 which shows the ratio of the male and female PI distributions. There is an increase in the number of early career (less than around 5 years from obtaining a PhD) female PIs compared to males; beyond ∼10 years after PhD, the ratio of females to all proposers is ∼20%, compared to ∼40% for the early career years.

We next investigated the acceptance rates for proposals from all proposers (male and female PIs) against PhD year. This is shown plotted for years −4 to 35 years after PhD in Fig. 11. A linear fit to the data gives an intercept of (37.3 ± 0.4)% and a gradient of (0.35 ± 0.15)% year$^{-1}$ assuming that the uncertainties are the square root of the number of people in each bin. The value of $R^2$ is 0.35. This is consistent with an increase from a ∼35% acceptance rate for PhD students to ~45% for senior researchers. Such an increase could result from a combination of factors including:

1. An increase in the success rate of researchers as they gain experience and have expanded networks of collaborators.
2. Less successful proposers who leave astronomy.
3. Less successful proposers who remain in astronomy, but do not propose in subsequent XMM-Newton AOs.
4. A bias in the selection process towards late career scientists.

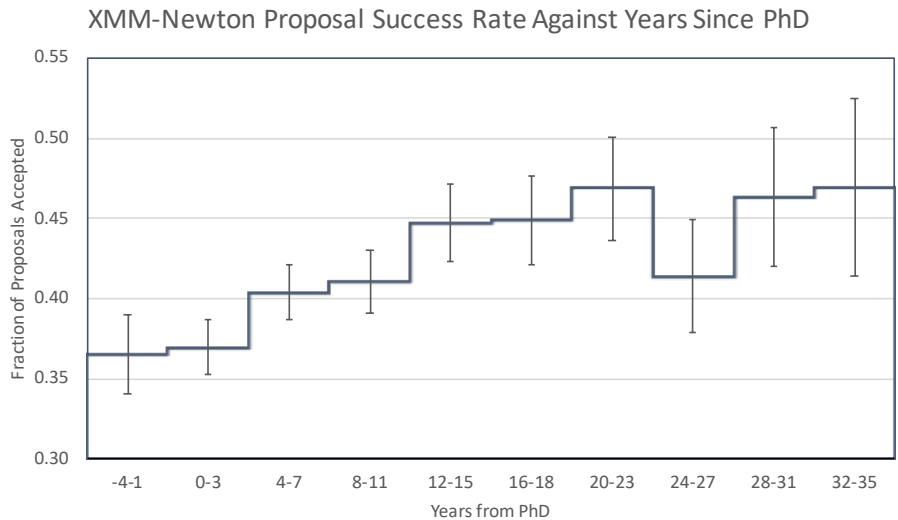

**Fig. 11** The acceptance fraction of XMM-Newton PIs with year of their PhD between −4 and 35 years after PhD. The indicative error bars indicate $1\sigma$ standard deviations assuming that the number of proposers in each age bin follows a Poisson distribution.

We note that the measurements are also consistent with an acceptance rate that remains approximately constant beyond ~12 years post-PhD and decreases more strongly for lower "academic ages".

The positive gradient of this relation indicates that the female proposer population is less likely to have proposals accepted simply due to having less experience. To calculate the size of this effect we took the male and female PI distributions shown in Fig. 9 and multiplied the number of PIs each year by the linear value derived from the fit to the overall success rate. This gave a predicted difference in the acceptance ratio of 1.0% between male and female PIs. This indicates that the underlying "academic age" differences between the two populations is probably not responsible for the majority of the differences in acceptance rate.

The acceptance fractions of male and female XMM-Newton PIs with year of their PhD are shown in Fig. 12. Proposals from male PIs have higher success rates in seven out of the eight bins. The success rate of proposals from male PIs is consistent with a gradual increase with "academic age" throughout the range examined (−4 to 35 years from PhD) from around 0.38 to nearly 0.50. For proposals with female PIs the success rate is consistent with increasing in a similar manner to their male counterparts until ~20 years post PhD, after which the gradual increase appears to stop and may reverse. We note that a similar decrease for senior female PIs is evident in HST Cycles 11 to 21 data [68]. We cannot exclude that proposals from female PIs simply have an average acceptance rate of ∼0.39, independent of "academic age".

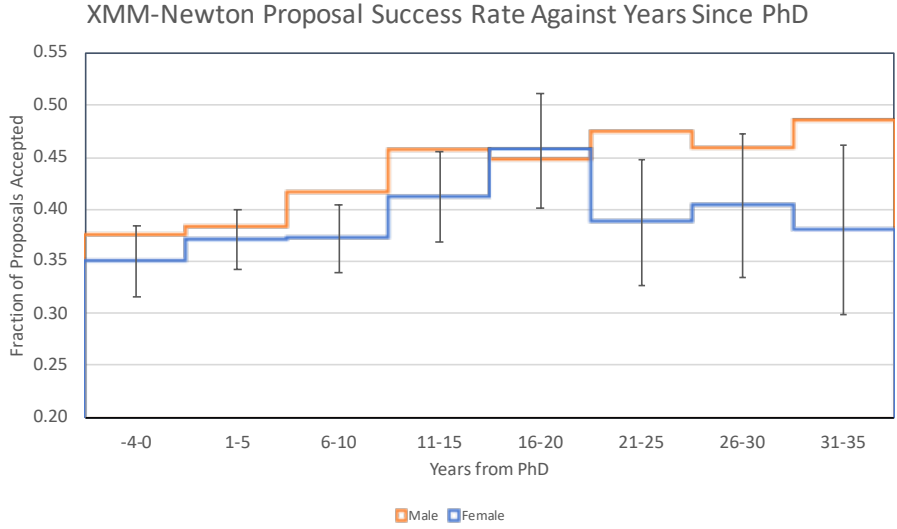

**Fig. 12** The acceptance fractions of male and female XMM-Newton PIs with year of their PhD between −4 and 35 years after PhD. The success rate of male PIs shows a gradual increase with "academic age" from ∼0.38 to nearly 0.50 over the age range considered. For female PIs the success rate may increase in a similar manner to their male colleagues until ∼20 years after PhD, after which the increase appears to stop and may even reverse. The indicative error bars indicate $1\sigma$ standard deviations assuming that the number of proposers in each age bin follows a Poisson distribution. Error bars are only shown for female PIs as these are much larger than for male PIs due to the smaller number of female PIs.

## 8.2 Proposal Selection – Priority A or B

We then investigated the outcomes of the proposals that were accepted with Priority A or B. Observations from these proposals are guaranteed to be performed, so having an approved high-priority proposal is likely to result in scientific papers and so be beneficial for the career of a researcher. This allows the investigation of any gender dependence of acceptance with assigned proposal priority. We note that an average of 42% of Priority C proposal targets were observed. Of the total of 10,579 proposals, 2467 were awarded observing time in priority A or B corresponding to an acceptance rate of 23.3%. Of the total of 7997 male-led and 2582 female-led proposals that were submitted, 1929 and 538 proposals were awarded observing time in priority A or B. This gives success rates of 24.1% and 20.9% for proposals with male and female PIs, respectively. This is a difference in favour of males of 15.6%. If we assume square root errors on the numbers of submitted and accepted proposals this implies success rates of $(24.1 \pm 0.6)\%$ and $(20.9 \pm 1.0)\%$ for proposals with male and female PIs, respectively and a difference in favour of males of $(15.6 \pm 6.2)\%$ which is significant at $2.5\sigma$. Square root uncertainties are unlikely to apply to the selection process and are only used to illustrate outcomes should this be the case.

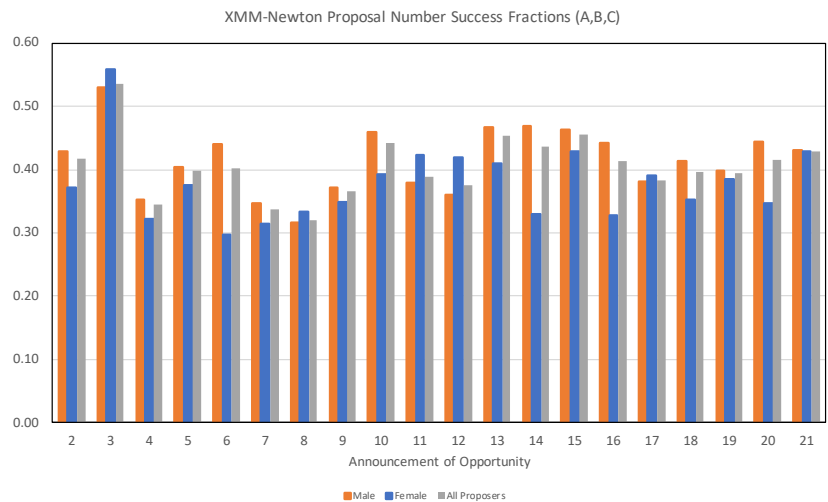

**Fig. 13** The fraction of accepted (Priority A, B or C) proposals compared to the number submitted. The histograms show the acceptance fractions for male PIs (orange), female PIs (blue) and all PIs (grey).

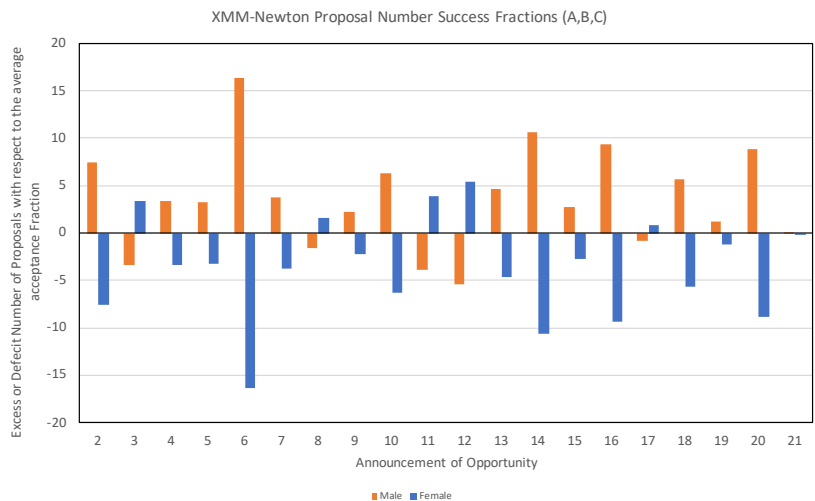

**Fig. 14** The relative success rates of XMM-Newton priority A, B and C proposals. The histograms show the difference between the actual number of successful proposals and the expected number based on the overall acceptance rate.

This difference is higher than when Priority C time is included. This implies that the OTAC approves relatively more Priority C proposals with female PIs compared to males PIs. The success rates are shown for male, female and all proposers in Fig. 18. Again, there is no obvious evolution in the male and female PI proposal acceptance rates with AO number, which cover an interval of more than 20 years. Figure 19 shows the variation in high-priority acceptance rates more clearly. For each AO, it shows the difference between the expected number of accepted proposals, calculated using the overall acceptance rate, and the number actually accepted. It shows that the largest discrepancies occurred during AO-6 and AO-16 where proposals with a female PI had lower success rates of 19.9% and 7.9%, respectively, compared to proposals with a male PI. The most successful AO for proposals with a female PI was AO-3 where such proposals were 9.5% more likely to be accepted than ones with a male PI. Figure 20 shows the same excesses normalised by dividing by the square root of the expected number of accepted proposals.

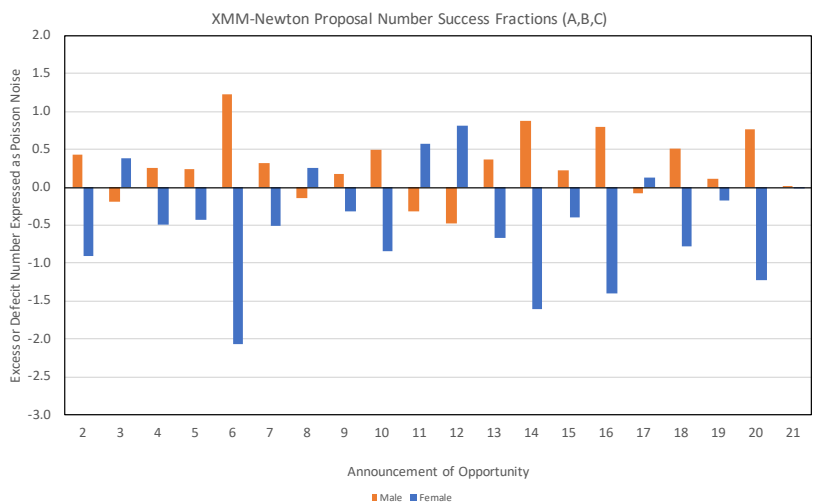

**Fig. 15** The relative success rates of XMM-Newton priority A, B and C proposals. The histograms show the difference between the actual number of successful proposals and the expected number based on the overall acceptance rate normalised by dividing by the square root of the number of expected proposals.

The average success rates for high priority (A or B) time is 14.6% of the requested times for male PIs and 12.6% for female PIs. This is a difference of 15.4% in favour of male proposers, which is very similar to the difference in the number of accepted high-priority proposals. This indicates that the assignment of high-priority observing time is handled similarly for proposals with male and female PIs.

### 8.3 Proposal Selection – Outcomes

We have examined the outcomes of 10,579 proposals submitted to ESA in response to the AO-2 to AO-21 calls for XMM-Newton observing time which were released between 2001 and 2021. The requested over-subscription in observing time was high ($\gtrsim$5) during all the AOs. During this time there was an increase in the fraction of female OTAC panel members from ~15% to ~25% of the total. The increase in female panel chairs is more marked from ~10% to ~45% of the total chairs by AO-21. Similarly, the fraction of female XUG members increased from ~5% to ~40% of the total. There is no obvious correlation between female PI success rates and the fraction of female OTAC members or panel chairs. We note that the lowest female success rate (0.143 during AO-6 for Priority A, B and C) occurred in one of only two AOs where the OTAC did not have any female panel chairs.

Scientists located at institutes within the USA submitted the most proposals – 40% of the total followed by those located in Germany, Italy and the UK. The Netherlands has the highest fraction of accepted proposals amongst the 13 countries submitting >80 proposals. Amongst these countries, Canada and Belgium have the highest fractions of proposals from female PIs; 50.3% and 42.7%, respectively. China and Switzerland have the lowest fractions; 11.2% and 16.5%, respectively. Three countries have better female PI proposal acceptance rates than for male PIs – the Netherlands, Japan and the United Kingdom with 7.3%, 4.7% and 1.3%, respectively compared to their averages.

Table 4 XMM-Newton AO proposal numbers showing for proposals with male and female PIs and the number of proposals accepted with high-priority (Priority A and B) and for all accepted (Priority A, B and C).

| AO | Submitted | | | Accepted | | | |
|---|---|---|---|---|---|---|---|
|  | No. of Proposals | Male led Proposals | Female led Proposals | Priority A+B Male | Priority A+B Female | Priority A+B+C Male | Priority A+B+C Female |
| 2 | 869 | 705 | 164 | 181 | 40 | 302 | 61 |
| 3 | 692 | 549 | 143 | 196 | 63 | 291 | 80 |
| 4 | 660 | 517 | 143 | 146 | 32 | 182 | 46 |
| 5 | 641 | 497 | 144 | 135 | 33 | 210 | 54 |
| 6 | 594 | 439 | 155 | 127 | 18 | 193 | 46 |
| 7 | 586 | 421 | 165 | 97 | 32 | 146 | 52 |
| 8 | 555 | 426 | 129 | 88 | 23 | 135 | 43 |
| 9 | 539 | 407 | 132 | 91 | 28 | 151 | 46 |
| 10 | 491 | 366 | 125 | 90 | 26 | 168 | 49 |
| 11 | 501 | 385 | 116 | 74 | 23 | 146 | 49 |
| 12 | 475 | 356 | 119 | 62 | 24 | 128 | 50 |
| 13 | 452 | 346 | 106 | 87 | 21 | 162 | 43 |
| 14 | 431 | 331 | 100 | 78 | 19 | 155 | 33 |
| 15 | 431 | 326 | 105 | 71 | 22 | 151 | 45 |
| 16 | 442 | 335 | 107 | 86 | 17 | 148 | 35 |
| 17 | 441 | 331 | 110 | 67 | 20 | 126 | 43 |
| 18 | 442 | 309 | 133 | 64 | 25 | 128 | 47 |
| 19 | 462 | 331 | 131 | 63 | 26 | 132 | 50 |
| 20 | 447 | 320 | 127 | 71 | 22 | 142 | 44 |
| 21 | 428 | 300 | 128 | 55 | 24 | 129 | 55 |
| **Total** | **10579** | **7997** | **2582** | **1929** | **538** | **3316** | **971** |

Table 5 XMM-Newton requested and accepted observing time for proposals with female and male PIs between AO-2 and AO-21 for high priority (A or B) and Priority A, B, and C. Time is in units of Msec.

| AO | Priority A+B | | | | | | Priority A+B+C | | | |
|---|---|---|---|---|---|---|---|---|---|---|
|  | Req. Time | | Acc. Time | | Success Rate | | Acc. Time | | Success Rate | |
|  | Male | Female | Male | Female | Male | Female | Male | Female | Male | Female |
| 2 | 96.5 | 21.7 | 12.8 | 3.0 | 0.132 | 0.136 | 21.5 | 4.6 | 0.223 | 0.212 |
| 3 | 81.9 | 16.9 | 15.5 | 3.9 | 0.190 | 0.231 | 26.8 | 6.0 | 0.327 | 0.353 |
| 4 | 80.8 | 21.1 | 11.8 | 3.1 | 0.146 | 0.145 | 15.2 | 4.1 | 0.187 | 0.196 |
| 5 | 87.3 | 19.5 | 9.6 | 3.1 | 0.110 | 0.159 | 16.3 | 4.8 | 0.187 | 0.245 |
| 6 | 77.0 | 23.1 | 11.7 | 1.2 | 0.152 | 0.051 | 17.3 | 3.3 | 0.225 | 0.143 |
| 7 | 87.8 | 26.0 | 9.6 | 3.5 | 0.109 | 0.136 | 14.2 | 6.8 | 0.161 | 0.263 |
| 8 | 96.6 | 21.5 | 11.4 | 1.7 | 0.118 | 0.077 | 17.6 | 4.3 | 0.182 | 0.199 |
| 9 | 87.7 | 26.5 | 10.2 | 2.6 | 0.117 | 0.100 | 16.1 | 4.8 | 0.184 | 0.180 |
| 10 | 66.6 | 22.9 | 9.1 | 5.3 | 0.137 | 0.230 | 17.3 | 7.7 | 0.259 | 0.338 |
| 11 | 66.4 | 21.1 | 9.4 | 2.5 | 0.142 | 0.118 | 15.6 | 6.0 | 0.235 | 0.286 |
| 12 | 61.5 | 24.4 | 9.5 | 2.4 | 0.154 | 0.098 | 15.8 | 5.0 | 0.257 | 0.204 |
| 13 | 58.9 | 18.8 | 10.8 | 3.7 | 0.183 | 0.197 | 18.2 | 6.1 | 0.309 | 0.325 |
| 14 | 60.8 | 24.1 | 11.7 | 2.2 | 0.193 | 0.103 | 21.3 | 3.7 | 0.350 | 0.173 |
| 15 | 56.5 | 24.1 | 10.1 | 3.1 | 0.179 | 0.129 | 19.0 | 5.4 | 0.337 | 0.223 |
| 16 | 65.9 | 25.2 | 12.7 | 2.1 | 0.193 | 0.085 | 21.1 | 4.0 | 0.320 | 0.158 |
| 17 | 96.3 | 38.5 | 11.6 | 5.3 | 0.120 | 0.137 | 20.0 | 7.7 | 0.208 | 0.199 |
| 18 | 63.7 | 29.3 | 8.0 | 3.9 | 0.125 | 0.132 | 16.2 | 9.4 | 0.254 | 0.320 |
| 19 | 66.8 | 25.1 | 7.9 | 2.9 | 0.118 | 0.117 | 17.6 | 6.7 | 0.264 | 0.268 |
| 20 | 76.1 | 30.6 | 15.4 | 2.3 | 0.202 | 0.074 | 24.3 | 6.1 | 0.319 | 0.198 |
| 21 | 56.7 | 24.3 | 8.5 | 3.1 | 0.151 | 0.127 | 17.0 | 7.5 | 0.300 | 0.307 |
| **Total** | **1491.8** | **481.6** | **217.3** | **60.8** |  |  | **368.4** | **113.8** |  |  |

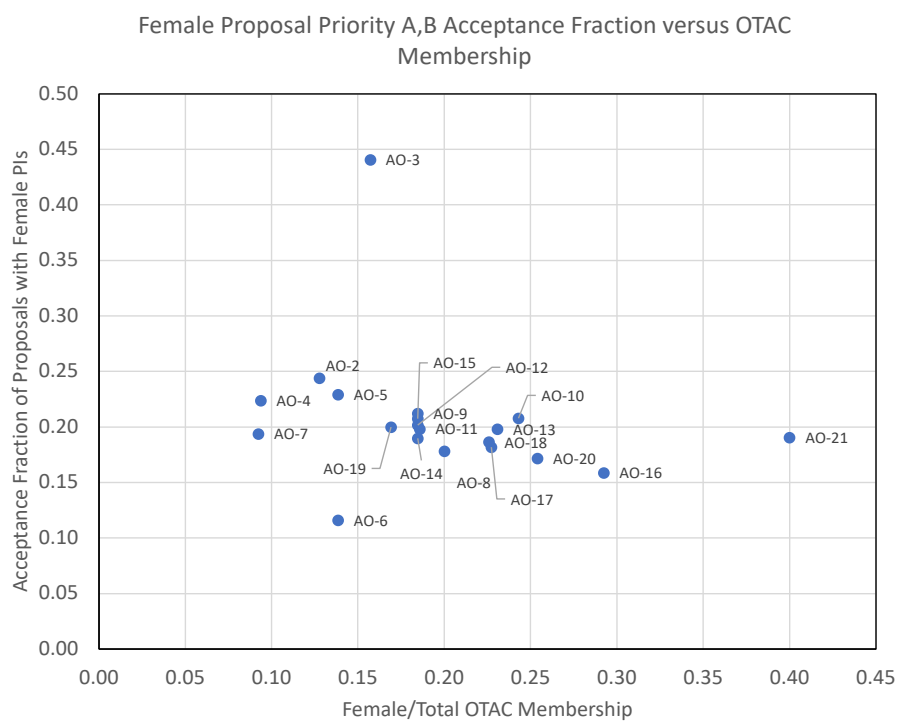

**Fig. 16** The acceptance fraction for female PIs for each AO (labelled) against the total fraction of female OTAC members. There is no obvious correlation between the two.

The fraction of XMM-Newton proposals from female PIs increased from ~20% of the total during the early AOs to ~30% of the total in the latest AOs. 41.5% of proposals with male PIs and 37.7% with female PIs were awarded any observing time on XMM-Newton. This is a difference in favour of males of 10.0%. There is no marked evolution in gender acceptance rates with AO number. The difference in success rates with gender is more marked when only proposals that were ranked to be high-priority (Priority A and B) are considered. These proposals are guaranteed to have their approved targets observed. The acceptance rate is 24.1% for proposals with male PIs and 20.9% for those with female PIs. This is a difference of 15.6%. This suggests that the OTAC ranks relatively more proposals with female PIs with a lower Priority C, than proposals with male PIs.

In order to investigate whether the OTAC awarded observing time differently for the two genders, the amounts of observing time awarded were also examined. For the high-priority proposals (priority A and B) 14.6% of the requested time was awarded for proposals with a male PI and 12.6% for proposals with a female PI. This is a difference of 15.4% in favour of male PIs. Similarly, for all the proposals, 24.7% of the requested time was awarded for proposals with a male PI and 23.6% for female PIs. The difference is 4.5% in favour of male PIs, smaller than the 10% difference when proposal numbers are considered. Again this implies that female PIs are relatively more likely to be awarded Priority C observing time. These results are summarised in Table 7. We interpret the different male/female PI proposal acceptance rates to imply that male PIs are between 5–15% more likely to benefit in the XMM-Newton proposal assessment process that their female counterparts.

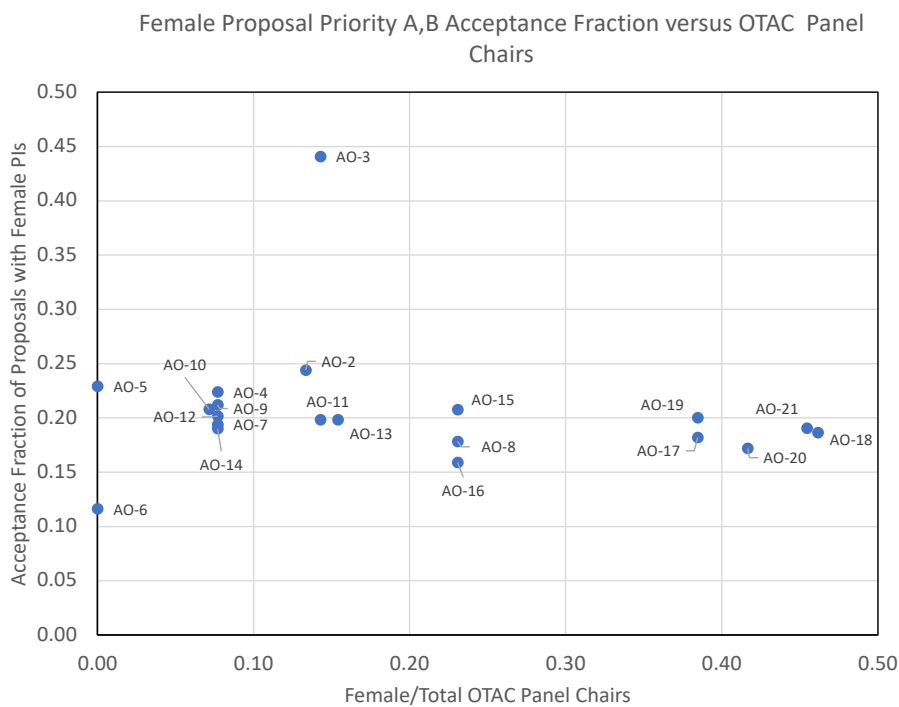

**Fig. 17** The acceptance fraction for female PIs for each AO (labelled) against the fraction of female OTAC panel chairs. There is no obvious correlation between the two.

Using the year of obtaining a PhD, or equivalent degree gives the mean "academic age" of PIs as 10.9 years post-PhD. For female PIs this is 9.8 years post-PhD compared to 11.2 years for male PIs. A linear fit to the proposal (Priority A, B and C) number acceptance rate against PhD year shows an increase in acceptance percentage from ∼35% prior to PhD to ∼45% for more "senior" astronomers. This increase in proposal acceptance probability with "academic age" favours a senior male population. However, this effect only accounts for ∼1% of the 5–15% difference in acceptance rates between male and female PIs.

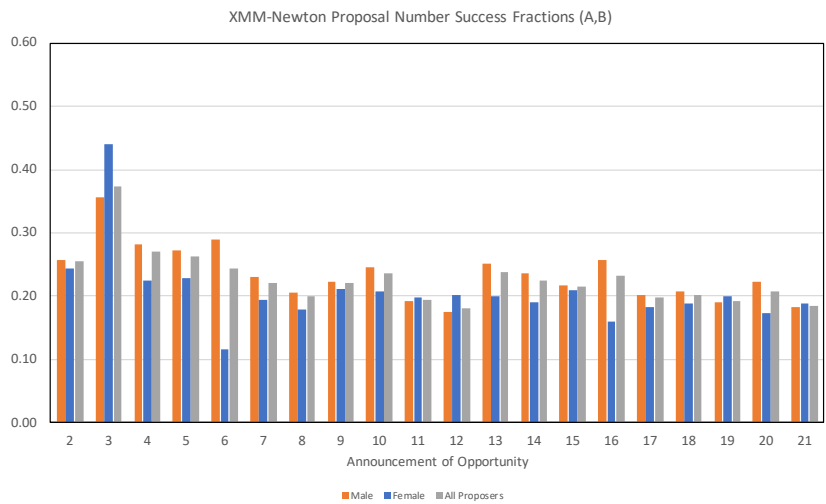

**Fig. 18** The fraction of high-priority (A or B) proposals compared to the number submitted. The histograms show the success rates for male PI (orange), female PI (blue) and all PIs (grey).

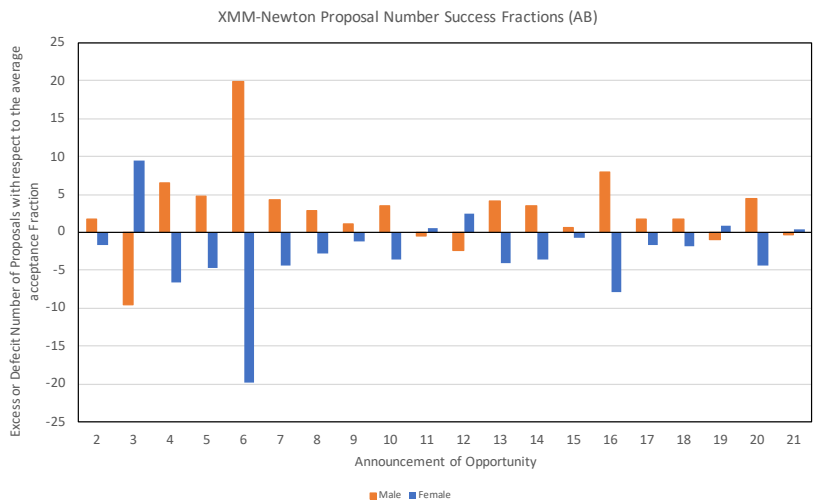

**Fig. 19** The relative success rates of XMM-Newton high-priority proposals. The histograms show the difference between the actual number of successful proposals and the expected number based on the average acceptance rates.

### 8.4 Proposal Selection – Regional Dependence

We have investigated the regional dependence of the gender outcomes observed above. We selected the proposals depending on the location of the institute of the PI. We used Europe, North America and the rest of the World as regions in a similar manner to [68] for HST. The number of proposals from each region and the fraction that were accepted are reported in Table 6. It is noticeable that proposals from Europe have the highest fraction of female

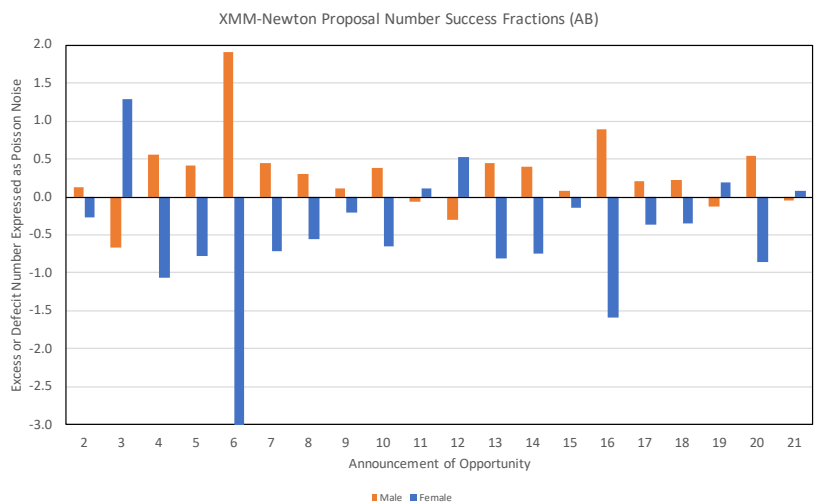

**Fig. 20** The relative success rates of high-priority XMM-Newton proposals. The histograms show the difference between the actual number of successful proposals and the expected number based on the overall acceptance rate normalised by dividing by the square root of the number of expected proposals.

**Table 6** The regional dependence of proposal submission and acceptance.

| Region | No. of Submitted Proposals | | | Total % | Female/Total % |
|---|---|---|---|---|---|
| | Total | Male | Female | | |
| Europe | 5277 | 3856 | 1421 | 49.9 | 26.9 |
| North America | 4467 | 3480 | 987 | 42.2 | 22.1 |
| Rest of the World | 835 | 661 | 174 | 7.9 | 20.8 |
| | No. of Accepted Proposals | | | | |
| | Total | Male | Female | | |
| Europe | 2270 | 1708 | 562 | | |
| North America | 1764 | 1407 | 357 | | |
| Rest of the World | 253 | 201 | 52 | | |
| | % Accepted Proposals | | | % Difference | |
| | Total | Male | Female | Male | Female |
| Europe | 43.0 | 44.3 | 39.5 | 1.3 | −3.5 |
| North America | 39.5 | 40.4 | 36.3 | 0.9 | −3.3 |
| Rest of the World | 30.3 | 30.4 | 29.9 | 0.1 | −0.4 |

PIs and overall success rates. Both the proposals from Europe and North America have higher success rates for male PIs compared to female PIs. The success rate for the rest of the World is lower than elsewhere with little difference between the male and female PI success rates. This is similar to the reported by [70] for Atacama Large Millimeter/sub-millimeter Array (ALMA).

## 9 Discussion

We have examined how our results compare to those from missions and facilities. HST is the obvious mission for comparison as it is similarly long-lived as XMM-Newton and has an even higher average number of proposals in response to each AO. There have also been investigations of the outcomes of the observing selection processes for ESO, Canadian facilities and the National Radio Astronomy Observatory (NRAO) facilities published, which we also consider.

There have been a number of other studies that have examined the gender dependence of AO outcomes. [68] reports on a study of gender-based systematic trends in the HST proposal review process for Cycles 11 through 21 (2001 to 2013). Key results from this study of the outcomes of 9400 proposals are that there appears to be a similar trend to that seen on XMM-Newton in that male PIs are more likely to succeed in achieving a successful HST proposal (23.5% success rate) than female PIs (18.1% success rate). This is a difference of 30% in favour of proposals led by male PIs. The effect is not necessarily significant in a single cycle but the imbalance is systematic with the success rate of HST female PIs consistently falling below that of their male colleagues. To help combat this effect, the Space Telescope Science Institute (STScI) adopted a system of dual-anonymous review, in which the names of the reviewers and the investigators are made known to each other only after the review has been completed. In HST Cycle 26, for the first time, proposals with female PIs had a higher success rate (8.7%) than those led by men (8.0%) [71]. However, this result is based on a single cycle which was unusual in that it had many fewer proposals than the average. Subsequent HST Cycles 27 to 29 (2019–2021) which were also conducted with dual-anonymous reviewing again showed higher male PI success rates. [72] report average male and female PI success rates of 16.2% and 14.8%, respectively for HST Cycles 26–29 which is a difference in favour of male PIs of 9.3%. The previous HST Cycles 22–25 had average male and female success rates of 25.1% and 22.2%. This is a larger difference than for Cycles 26–29 in favour of male PIs of 13.2%. Thus in HST Cycles 27 to 29 dual-anonymous reviewing probably *helped* the HST AO selection process to provide more equitable outcomes. In addition, as [72] reports, the HST success rate of early career scientists increased under dual-anonymous reviewing from ∼5% to ∼30% with a significant increase in the number of first-time PIs. It is unclear if this leads to a more scientifically productive mission.

The analysis presented here indicates that the XMM-Newton proposal selection process has comparable gender outcomes to that of HST after dual-anonymous reviewing was implemented by STScI. Note that we do not refer to a "bias" in the XMM-Newton proposal selection since the different success rates may result from factors not associated with the proposal selection process itself. The XMM-Newton proposal selection process is clearly well optimised, as would be expected following so many AO calls. We do not see a strong case for the implementation of dual-anonymous proposal handling for XMM-Newton. Such a change could have unexpected consequences given the complexities of the selection process.

An investigation of the time allocation process at ESO is reported in [73]. This covers an interval of 8 years and involved about 3000 PIs. Female PIs were found to have a significantly lower chance of being awarded a top rank compared to male PIs with a male/female ratio of $1.39 \pm 0.05$. The paper suggests that the principal explanation for the difference may be due to the average higher seniority of the male PIs assuming that more senior scientists of both genders write better proposals and thus succeed more often at obtaining ESO telescope time. However, no attempt was made to quantify if this effect can account for the observed differences. Nevertheless, [73] concludes that the ESO review process itself introduces extra gender differences.

The results of the Canadian Time Allocation Committee rankings for the Canadian share of the Canada-France-Hawaii Telescope (CFHT) and Gemini Observatory time have been investigated by [74]. They find that proposals led by males are more likely to obtain higher scores than those led by females. This difference was present for both proposers who had and did not have faculty positions implying that seniority was not the cause. An investigation of the gender-related systematics in the proposal review processes for the four facilities operated by NRAO: the Jansky Very Large Array (JVLA), the Very Long Baseline Array (VLBA), the Green Bank Telescope (GBT) and ALMA are reported in [75] and [70]. Similarly to HST, before the introduction of dual-anonymous reviewing, and similarly as well to the ESO study there are significant gender-related effects in the proposal rankings in favour of male PIs compared to female PIs for all four facilities with varying degrees of confidence reflecting the different number of proposals.

**Table 7** Summary of XMM-Newton proposal acceptance percentages for AO-2 to AO-21 for both proposal numbers and awarded time. The overall male/female difference covers the range of given male/female ratios and is expressed as a percentage.

| Priority | Parameter | Proposal PI | | |
|---|---|---|---|---|
| | | All | Male | Female |
| | Proposals Submitted | 10,579 | 7997 | 2582 |
| A,B,C | Number Accepted | 4287 | 3316 | 971 |
| | Percentage Accepted | 40.5% | 41.5 | 37.7 |
| | Ratio Male/Female | | 1.100 | |
| A,B | Number Accepted | 2467 | 1929 | 538 |
| | Percentage Accepted | 23.3 | 24.1 | 20.9 |
| | Ratio Male/Female | | 1.155 | |
| | Time Requested (s) | $1.97 \times 10^9$ | $1.49 \times 10^9$ | $4.81 \times 10^8$ |
| A,B,C | Time Accepted (s) | $4.82 \times 10^8$ | $3.68 \times 10^8$ | $1.14 \times 10^8$ |
| | Percentage Time Accepted | 24.4% | 24.7% | 23.6% |
| | Ratio Male/Female | | 1.045 | |
| A,B | Time Accepted(s) | $2.78 \times 10^8$ | $2.17 \times 10^8$ | $6.08 \times 10^7$ |
| | Percentage Time Accepted | 14.1% | 14.6% | 12.6% |
| | Ratio Male/Female | | 1.154 | |
| | Overall Male/Female Difference | | 5–15% | |

In summary, the difference observed in the XMM-Newton proposal selection process of 5–15% in favour of male PIs is smaller than those reported for some ESO, and NRAO facilities [73], [75] and for HST prior to the implementation of dual-anonymous reviewing [68]. Once dual-anonymous reviewing was implemented for HST the male and female PI proposal acceptance difference falls to 9.3% [72], comparable to that seen here with XMM-Newton of 5–15%. The success rate of scientists prior to obtaining their PhDs of ∼35% on XMM-Newton is comparable to that on HST following dual-anonymous reviewing. This suggests that the XMM-Newton proposal selection processes produces outcomes consistent with, or at least close to, the best available science. It does not however provide gender parity and a number of activities are proposed below to investigate this matter.

## 10 Further Activities

In order to better understand some of the issues arising from the analysis of the XMM-Newton proposal selection processes, the following would be useful:

1. An investigation of the uncertainties in the proposal selection process. This could be done by comparing the results of different representative OTACs evaluating the same

    set of proposals. Given the amount of work involved in an OTAC evaluation this would be a major undertaking. Instead, it may be possible to gain insights into the process by looking at how proposals that are evaluated by multiple panels are assessed.
2. A study of the impact of native language on the success rate of proposals. Non-native language is increasingly being recognised as a major barrier in science [76] implying that native English speakers are more successful than their counterparts in obtaining observing time. This does not appear to be the case for XMM-Newton as countries which are expected to have large fractions of native English speakers, such as the United Kingdom and the United States, do not have unusually high success rates for their XMM-Newton proposals (see Table 2).
3. A further investigation of the acceptance rates of male and female proposers for different geographical regions – for example dividing Europe into Northern and Southern regions. This could show that regional differences play a role and provide insights into the causes of the different success rates.
4. An investigation into the publication rates of early and late career scientists who have been awarded XMM-Newton observing time. Are late career scientists better at exploiting their observations and so produce relatively more publications and citations than their early career colleagues? More generally, what is the effect on the science return of a mission if the observations are dominated by late or early career scientists. This is a complex issue as e.g., a late career scientist may have made substantial contributions to a proposal from an early career colleague, so improving its chances of success.

**Acknowledgements.** We thank Pedro Rodriquez and the XMM-Newton SOC staff. Fred Jansen is acknowledged for his efforts in astro-archeology. We thank Neill Reid at the STScI who provided the results of the most recent HST observing cycles.

## Acronym List

| | |
|---|---|
| **ADS** | Astrophysics Data Service |
| **AGN** | Active Galactic Nucleus |
| **AO** | Announcement of Opportunity |
| **ALMA** | Atacama Large Millimeter/sub-millimeter Array |
| **CFHT** | Canada-France-Hawaii Telescope |
| **EPIC** | European Photon Imaging Camera |
| **ESA** | European Space Agency |
| **ESO** | European Southern Observatory |
| **FOV** | Field of View |
| **FWHM** | Full-Width at Half Maximum |
| **GBT** | Green Bank Telescope |
| **GSFC** | Goddard Space Flight Center |
| **HESS** | High Energy Stereoscopic System |
| **HST** | Hubble Space Telescope |
| **JAXA** | Japanese Aerospace Exploration Agency |
| **JVLA** | Jansky Very Large Array |
| **keV** | Kilo Electron Volt |
| **NASA** | National Aeronautics and Space Administration |
| **NRAO** | National Radio Astronomy Observatory |
| **ObsID** | Observation Identifier |
| **OM** | Optical Monitor |
| **OTAC** | Observation Time Allocation Committee |
| **PhD** | Doctor of Philosophy Degree |
| **PI** | Principal Investigator |
| **RGS** | Reflection Grating Spectrometer |
| **SOC** | Science Operations Centre |
| **STScI** | Space Telescope Science Institute |
| **TDE** | Tidal Disruption Event |
| **ToO** | Target of Opportunity |
| **UV** | Ultra-Violet |
| **VLBA** | Very Long Baseline Array |
| **VLT** | Very Large Telescope |
| **WHIM** | Warm-Hot Intergalactic Medium |
| **XSA** | XMM-Newton Science Archive |
| **XUG** | XMM-Newton Users' Group |